\documentclass[aps,prd,preprint,nofootinbib]{revtex4}

\usepackage{graphicx}
\usepackage{psfrag}
\usepackage{epsfig}

\begin{document}

\title{A Comprehensive Analysis of the Pion-Photon Transition Form Factor Beyond the Leading Fock State}
\author{Tao Huang$^{1,2}$\footnote{email:
huangtao@mail.ihep.ac.cn} and Xing-Gang Wu$^{3}$\footnote{email:
wuxg@itp.ac.cn}}
\address{$^1$CCAST(World Laboratory), P.O.Box 8730, Beijing 100080,
P.R.China\\
$^2$Institute of High Energy Physics, Chinese Academy of Sciences,
P.O.Box 918(4), Beijing 100039, P.R. China\\
$^3$Institute of Theoretical Physics, Chinese Academy of Sciences,
P.O.Box 2735, Beijing 100080, P.R. China}

\date{\today}

\begin{abstract}
We perform a comprehensive analysis of the pion-photon transition
form factor $F_{\pi \gamma}(Q^2)$ involving the transverse momentum
corrections with the present CLEO experimental data, in which the
contributions beyond the leading Fock state have been taken into
consideration. As is well-known, the leading Fock-state contribution
dominates of $F_{\pi \gamma}(Q^2)$ at large momentum transfer
($Q^2$) region. One should include the contributions beyond the
leading Fock state in small $Q^2$ region. In this paper, we
construct a phenomenological expression to estimate the
contributions beyond the leading Fock state based on its asymptotic
behavior at $Q^2\to0$. Our present theoretical results agree well
with the experimental data in the whole $Q^2$ region. Then, we
extract some useful information of the pionic leading twist-2
distribution amplitude (DA) by comparing our results of $F_{\pi
\gamma}(Q^2)$ with the CLEO data. By taking best fit, we have the DA
moments, $a_2(\mu_0^2)=0.002^{+0.063}_{-0.054}$,
$a_4(\mu_0^2)=-0.022_{-0.012}^{+0.026}$ and all of higher moments,
which are closed to the asymptotic-like behavior of the pion
wavefunction. \\


\end{abstract}

\maketitle

\section{Introduction}

The pion-photon transition form factor, which relates two photons
with one lightest meson, is the simplest example for the
perturbative application to exclusive processes. By neglecting
$k_\bot$ (the transverse momentum of the constitute quarks) relative
to $q_\bot$ (the transverse momentum of the virtual photon) in the
hard-scattering amplitude, one can obtain the leading Fock-state
formula \cite{lb}:
\begin{eqnarray}
F_{\pi \gamma} (Q^2)=\frac{2f_\pi}{3 Q^2} \int \frac{dx}{x}
\phi_\pi(x,Q^2) \left[ 1+{\cal O}\left(\alpha_s(Q^2),
\frac{m^2}{Q^2}\right) \right], \label{fLepage}
\end{eqnarray}
where $Q^2=-q^2$ stands for the momentum transfer in the process,
the pion decay constant $f_{\pi}=92.4\pm0.25$~MeV~\cite{pdg}, and
$\phi_\pi(x,Q^2)$ stands for the leading Fock-state pion
distribution amplitude (DA) at the factorization scale $\mu^2=Q^2$,
which can be derived from the initial DA $\phi_\pi(x,\mu_0^2)$
($\mu_0$ stands for some hadronic scale) through QCD evolution. The
initial DA is defined as
\begin{equation}\label{DAdefinition}
\phi_\pi(x,\mu_0^2)=\frac{2\sqrt{3}}{f_\pi}
\int_{|\mathbf{k}_\perp|^2\leq\mu_0^2}\frac{d^2\mathbf{k}_\perp}{16\pi^3}
\Psi_{q\bar{q}}(x,\mathbf{k}_\perp).
\end{equation}
Hence, the value of $Q^2 F_{\pi \gamma} (Q^2)$ tends to be a
constant ($2f_\pi$) for asymptotic DA:
$\phi_{as}(x,Q^2)|_{Q^2\rightarrow \infty}=6x(1-x)$. However, it has
been argued that the $k_\perp$-dependence of the pion wavefunction
can not be safely neglected at the end-point region $x_i \rightarrow
0,1$ ($x_i$ the momentum fraction of the constitute quarks in pion)
and $Q^2 \sim$ a few GeV$^2$. Under the light-cone (LC) pQCD
approach, the leading contribution to $F_{\pi \gamma}(Q^2)$ that
keeps the $k_\bot$-corrections in both the hard-scattering amplitude
and the wavefunction can be written as \cite{bhl,huang1}:
\begin{equation}
F_{\pi \gamma}(Q^2)=2 \sqrt{3}(e_u^2-e_d^2)\int_0^1 [dx]\int
\frac{{\rm d}^2 \mathbf{k}_\perp}{16 \pi^3}
\Psi_{q\bar{q}}(x,\mathbf{k}_\perp) \times
T_H(x,x^{\prime},\mathbf{k}_\perp),\label{fflead}
\end{equation}
where $[dx]=dxdx'\delta(1-x-x')$, $e_{u,\; d}$ are the quark charges
in unites of $e$, $\Psi_{q\bar{q}}(x,\mathbf{k}_\perp)$ stands for
the leading Fock-state wavefunction and the hard-scattering
amplitude $T_H(x,x',\mathbf{k}_\perp)$ takes the form,
\begin{equation}\label{hardlead}
T_H(x,x',\mathbf{k}_\perp)= \frac{\mathbf{q}_\perp \cdot (x'
\mathbf{q}_\perp + \mathbf{k}_\perp)} {\mathbf{q}_\perp^2 (x'
\mathbf{q}_\perp + \mathbf{k}_\perp)^2} +(x \leftrightarrow x').
\end{equation}
With the help of Eqs.(\ref{fflead},\ref{hardlead}),
Ref.\cite{huang1} performed a careful analysis of the quark
transverse-momentum effect to $F_{\pi \gamma}(Q^2)$. They pointed
out that the transverse-momentum dependence in both the numerator
and the denominator of the hard-scattering amplitude are of the same
importance and should be considered consistently. Similar improved
treatment has also been done in
Refs.\cite{rus1,job1,ma1,musatov,melic,kim,ong}. It was shown that
pQCD can give the correct prediction for the pion-photon transition
form factor that is consistent with the present experimental data by
keeping the $k_\perp$-dependence in both the hard-scattering
amplitude and the pion wavefunction and by properly choosing of the
pion wavefunction.

It should be noted that Eqs.(\ref{fLepage},\ref{fflead}) were
obtained by assuming the leading Fock-state dominance. This
approximation is valid only for large $Q^2$ region and one can not
expect that these expressions can describe the present experimental
data well in low $Q^2$ region. Refs.\cite{pqcd,pqcd2} show that the
approximation of the leading Fock-state dominance to the pion
electro-magnetic form factor is valid as $Q^2\gtrsim 4$~GeV$^2$,
which is improved to be $Q^2\gtrsim 1$~GeV$^2$ by including the
next-to-leading order (NLO) contribution \cite{yeh,a2a44}. A similar
discussion has been done in Ref.\cite{huang1} for the pion-photon
transition form factor. These references tell us that one should
take into account the higher Fock states' contributions as $Q^2<a
\;few$~GeV$^2$. In fact, it has been shown that the expression
(\ref{fflead}) gives half contribution to $F_{\pi\gamma}(Q^2)$ as
one extends it to $Q^2=0$~\cite{bhl}. It means that the leading Fock
state contributes to $F_{\pi\gamma}(0)$ only half and the remaining
half should be come from the higher Fock states. Both contributions
from the leading Fock state and the higher Fock states are needed to
get the correct $\pi^0\to\gamma\gamma$ rate~\cite{bhl}. Any attempt
that involves only the leading Fock-state contribution to explain
both the $\pi^0\to\gamma\gamma$ rate and the pion-photon transition
form factor for low $Q^2$ region is incorrect. It should be pointed
out that the above conclusion does not contradict with that of
Ref.\cite{rady2}, where with the help of an ``effective" two-body
wavefunction that includes the soft contributions from the higher
Fock components, the authors pointed that the contributions
corresponding to higher Fock states in a hard region appear as
radiative corrections and are suppressed by powers of
$(\alpha_s/\pi)\sim 10\%$.

In this paper, we will take the contributions from both the leading
Fock state and the higher Fock states into consideration.
Especially, we will discuss how to consider the contributions beyond
the leading Fock state at low $Q^2$ region and give a careful
analysis of the pion-photon transition form factor in the whole
$Q^2$ region. Furthermore, we can learn more information of the
leading twist-2 DA from the present CLEO data, since the pion-photon
transition form factor in the simplest exclusive process only
involves one pion and the contributions from the higher twist
structures and higher helicity states are highly suppressed (at
least by $1/Q^4$) in comparison to the leading twist-2 pion
wavefunction. In the present paper, we will not include the NLO
contribution into our formulae. Since we need a full NLO result in
order to be consistent with our present calculation technique, in
which the effects caused by the transverse-momentum dependence in
the hard-scattering amplitude and the pion wavefunction and by the
Sudakov factor should be fully considered, however such a full NLO
calculation is not available at the present \footnote{The present
NLO result is derived without considering the transverse-momentum
dependence in the hard-scattering amplitude and the wavefunction,
see e.g. Refs.\cite{brodsky2,melic,yeh,a2a44,musatov}.}.

The paper is organized as follows. In Sec.II, we analyze the
contributions to $F_{\pi \gamma}(Q^2)$ beyond the leading Fock state
at low $Q^2$ region under the LC pQCD approach and give a complete
expression for $F_{\pi \gamma}(Q^2)$ in the whole $Q^2$ region. In
Sec.III, we discuss what we can learn of the pionic leading
Fock-state wavefunction/DA in comparison with CLEO experimental
data. Some further discussion and comments are made in Sec.IV. The
last section is reserved for a summary.

\section{An expression of $F_{\pi \gamma}(Q^2)$ from zero to large $Q^2$ region}

First, we give a brief review of the LC formalism
~\cite{lb,lbhm,bpp}. The LC formalism provides a convenient
framework for the relativistic description of hadrons in terms of
quark and gluon degrees of freedom and for the application of pQCD
to exclusive processes. The LC Fock-state expansion of wavefunction
provides a precise definition of the parton model and a general
method to calculate the hadronic matrix element. As for the pion
wavefunction, its Fock-state expansion is
\begin{eqnarray}
| \pi \rangle=\sum | q \bar{q} \rangle \Psi_{q \bar{q}}+ \sum | q
\bar{q} g \rangle \Psi_{q \bar{q} g}+\cdots, \label{Fock}
\end{eqnarray}
where the Fock-state wavefunctions $\Psi_{n}(x_i,\mathbf{k}_{\perp
i},\lambda_i)$ $(n=2,3,\cdots)$ satisfy the normalization condition
\begin{equation}
\sum_{n,\lambda_i}\int
[dxd^2\mathbf{k}_\perp]_n|\Psi_{n}(x_i,\mathbf{k}_{\perp
i},\lambda_i)|^2=1 ,
\end{equation}
with $[dxd^2\mathbf{k}_\perp]_n=16\pi^3 \delta(1-\sum_{i=1}^n x_i)
\delta^2(\sum_{i=1}^n\mathbf{k}_{\perp
i})\prod_{i=1}^n\left[\frac{dx_i d^2\mathbf{k}_{\perp
i}}{16\pi^3}\right] $. $\lambda_i$ is the helicity of the
constituents and $n$ stands for all Fock states, e.g.
$\Psi_2=\Psi_{q\bar{q}}$. It should be pointed out that we have
$\int [dxd^2\mathbf{k}_\perp]_2 \sum_{\lambda_i}
|\Psi_{q\bar{q}}(x_i,\mathbf{k}_{\perp i},\lambda_i)|^2<1$ for the
leading Fock state.

The pion-photon transition form factor is connected with the $\pi^0
\gamma \gamma^*$ vertex in the amplitude of $e \pi \rightarrow e
\gamma$ as
\begin{eqnarray}
\Gamma_\mu =- i e^2 F_{\pi \gamma}(Q^2) \epsilon_{\mu \nu \alpha
\beta} P^\mu \epsilon^\alpha q^\beta,
\end{eqnarray}
where $P$ and $q$ are the momenta of the incident pion and the
virtual photon respectively, and $\epsilon$ is the polarization
vector of the final on-shell photon. To simplify the hard-scattering
amplitude, we adopt the Drell-Yan-West assignment at the
`infinite-momentum' frame~\cite{drell}: $
q=(q_+,q_-,\mathbf{q}_\perp)=(0,|\mathbf{q}_\perp|^2/P^+,\mathbf{q}_\perp)$
and $ P=(P^+,P^-,\mathbf{P}_\perp)=(1,0,\mathbf{0}_\perp)$, where
$P^+$ is arbitrary because of Lorentz invariance and it is taken to
be $1$ for convenience. Then $F_{\pi \gamma}$ is given by \cite{lb}
\begin{eqnarray}
F_{\pi \gamma}(Q^2)=\frac{\Gamma^+}{-i e (\mathbf{\epsilon}_\perp
\times \mathbf{q}_\perp)},
\end{eqnarray}
where $Q^2=-q^2=|\mathbf{q}_\perp|^2$,
$\epsilon=(0,0,\mathbf{\epsilon}_\perp)$, $\mathbf{\epsilon}_\perp
\cdot \mathbf{q}_\perp=0$ and $\mathbf{\epsilon}_\perp \times
\mathbf{q}_\perp=\epsilon_{\perp 1} q_{\perp 2}-\epsilon_{\perp 2}
q_{\perp 1}$.

\begin{figure}
\centering
\includegraphics[width=0.70\textwidth]{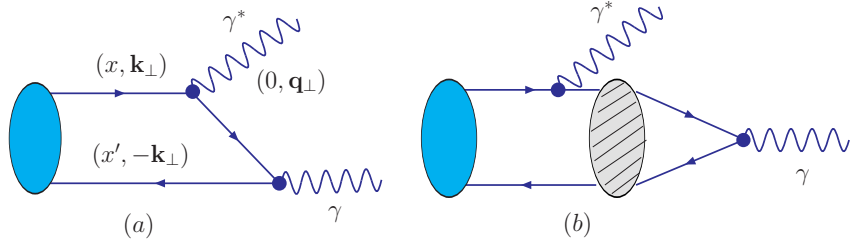}
\caption{Typical diagrams that contribute to the pion-photon
transition form factor $F_{\pi \gamma}(Q^2)$, where $x'=(1-x)$. The
rightmost shaded oval with a slant pattern stands for the strong
interactions.} \label{feyn}
\end{figure}

As illustrated in Fig.(\ref{feyn}), there are two basic types of
contribution to $F_{\pi \gamma}(Q^2)$ , i.e. $F^{(V)}_{\pi
\gamma}(Q^2)$ and $F^{(NV)}_{\pi \gamma}(Q^2)$. $F^{(V)}_{\pi
\gamma}(Q^2)$ comes from Fig.(\ref{feyn}a), which involves the
direct annihilation of $(q\bar{q})$-pair into two photons, i.e. the
leading Fock-state contribution that dominates the large $Q^2$
contribution. $F^{(NV)}_{\pi \gamma}(Q^2)$ comes from
Fig.(\ref{feyn}b), in which one photon coupling `inside' the pion
wavefunction, i.e. strong interactions occur between the photon
interactions that is related to the higher Fock states'
contributions. An interpretation for $F^{(NV)}_{\pi \gamma}(Q^2)$
can be given under the operator product expansion (OPE) approach
\cite{rady2}. Under the OPE approach \cite{ope}, the nonperturbative
aspects of the hadron dynamics are described by matrix elements of
local operators. In particular, the longitudinal momentum
distribution is related to the lowest-twist composite operators.
While by taking into account the transverse-momentum effects, one
needs to consider matrix elements of higher-twist composite
operators in which some of the covariant derivatives appear in a
contracted form like $D^2 = D_{\mu} D^{\mu}$. By using the equation
of motion for the light quark, $\gamma^{\mu} D_{\mu} q = 0$, one can
convert a two-body quark-antiquark operator $  \bar q \{
\gamma_{\mu_1} D_{\mu_2} \ldots D_{\mu_n} \} D^2 q$ into the
``three-body'' operator $  \bar q \{ \gamma_{\mu_1} D_{\mu_2} \ldots
D_{\mu_n} \} (\sigma^{\mu \nu} G_{\mu\nu}) q$ with an extra gluonic
field $G_{\mu\nu}$ being involved, which is just related to the
higher Fock state of pion.

The first type of contribution $F^{(V)}_{\pi \gamma}(Q^2)$ stands
for the conventional leading Fock-state contribution. Under the LC
pQCD approach and by keeping the full $k_T$-dependence in both the
hard-scattering amplitude and the wavefunction, the expression for
$F^{(V)}_{\pi \gamma}(Q^2)$ is the one that is given in
Eqs.(\ref{fflead},\ref{hardlead}), where the terms involving the
higher helicity states ($\lambda_1+\lambda_2=\pm1$, $\lambda_i$
stands for the corresponding helicity of the two constitute quarks
of pion) and the higher twist structures of the pion wavefunction
are not explicitly written. Since by direct calculating, one may
observe that the contributions from the higher helicity states and
the higher twist structures of the pion wavefunction are suppressed
by at least $1/Q^4$ to that of the usual helicity state
$(\lambda_1+\lambda_2=0)$ of the leading Fock state, which agrees
with the discussion made in Ref.\cite{musatov} \footnote{The present
condition is quite different from the case of pion electro-magnetic
form factor, where the contributions from the higher helicity states
and higher twist structures are only suppressed by $1/Q^2$ and then
they can provide sizable contribution in the intermediate $Q^2$
region \cite{highhelicity,hightwist}.}.

As for the second type of contribution $F^{(NV)}_{\pi \gamma}(Q^2)$,
it is difficult to be calculated in any $Q^2$ region. If treating
the photon vertex in Fig.(\ref{feyn}b) as a vector meson dressed
photon vertex, Fig.(\ref{feyn}b) can be calculated approximately
under the vector meson dominance (VMD) approach, see e.g.
Ref.\cite{vmd} for a review and Refs.\cite{vmd2,vmd3} for an
explicit VMD calculation of the pion electro-magnetic form factor.
By adopting the VMD approach to approximate Fig.(\ref{feyn}b), one
needs to introduce some undetermined coupling factor either for VMD1
or VMD2 formulation \cite{vmd}, which together with the undetermined
parameters in the pion wavefunction can not be definitely determined
by the CLEO experimental data of the pion transition form factor
only and some other constraints should also be taken into
consideration, e.g. the constraint from the experimental value for
the pion charge radius or the constraint from the experimental value
for the pion electro-magnetic form factor. In fact, one usually
takes the value of $F_{\pi \gamma}(Q^2)=F^{(V)}_{\pi\gamma}(Q^2)+
F^{(NV)}_{\pi\gamma}(Q^2)$ derived from the VMD approach to be in a
simple monopole form~\cite{monopole}, i.e. $F_{\pi
\gamma}(Q^2)=1/\left[4\pi^2 f_\pi(1+Q^2/m_\rho^2)\right]$, with the
$\rho$-meson mass $m_\rho$ serves as a parameter determined by the
pion charge radius. For the purpose of extracting some useful
information of the pion wavefunction from the CLEO experimental
data, we adopt the method raised by Ref.\cite{bhl} to deal with
$F^{(NV)}_{\pi \gamma}(Q^2)$ \footnote{A careful VMD calculation of
the pion transition form factor along the line of
Refs.\cite{vmd2,vmd3} can be used as a cross check of our results,
however it is out of the range of the present paper.}.

As stated in Ref.\cite{bhl}, around the region of $Q^2\sim 0$, since
the wavelength of the photon `inside' the pion wavefunction $\sim
1/m_{\pi}$ is assumed to be much larger than the pion radius
$1/\lambda$ ($\lambda$ is some typical hadronic scale $\sim 1$~GeV),
we can treat such photon (nearly on-shell) as an external field
which is approximately constant throughout the pion volume. And
then, a fermion in a constant external field is modified only by a
phase, i.e. $S_A(x-y)=e^{-ie(y-x)\cdot A}S_F(x-y)$. Consequently,
the lowest $q\bar{q}$-wavefunction for the pion is modified only by
a phase $e^{-i e y\cdot A}$, where $y$ is the $q\bar{q}$-separation.
Transforming such phase into the momentum space and applying it to
the wavefunction, the second contribution $F^{(NV)}_{\pi
\gamma}(Q^2)$ at $\mathbf{q}_\perp\to 0$ can be simplified to
\begin{equation}\label{ffb0}
F^{(NV)}_{\pi \gamma}(Q^2)|_{\mathbf{q}_\perp\to
0}=\frac{-2}{\sqrt{3}Q^2}\int [dx] \int
\frac{d^2\mathbf{k}_\perp}{16\pi^3}\left\{\frac{(\mathbf{k}_\perp
\times\mathbf{q}_\perp)^2} {(x'\mathbf{q}_\perp+
\mathbf{k}_\perp)^2}\left[ \frac{\partial}{\partial k_\perp^2}
\Psi_{q\bar{q}}(x,\mathbf{k}_{\perp}) \right] +(x\leftrightarrow
x')\right\},
\end{equation}
where $[dx]=dxdx'\delta(1-x-x')$ and $k_\perp=|\mathbf{k}_\perp|$.
Eq.(\ref{ffb0}) gives the expression for $F^{(NV)}_{\pi
\gamma}(Q^2)$ at $Q^2\to 0$. Here different from Ref.\cite{bhl}, all
$\mathbf{q}_\perp$-terms that are necessary to obtain its first
derivative over $Q^2$ are retained and the relation
$(\mathbf{\epsilon}_\perp\times\mathbf{q}_\perp)
(\mathbf{k}_\perp\times\mathbf{q}_\perp)=Q^2(\mathbf{\epsilon}_\perp
\cdot\mathbf{k}_\perp)$ is implicitly adopted. After doing the
integration over $\mathbf{k}_\perp$, one can easily find that
\begin{equation} \label{ffb00}
F^{(NV)}_{\pi \gamma}(0)=F^{(V)}_{\pi \gamma}(0)
=\frac{1}{8\sqrt{3}\pi^2}\int dx\Psi_{q\bar{q}}(x,\mathbf{0}_\perp),
\end{equation}
which means that the leading Fock state contributes to
$F_{\pi\gamma}(0)=F^{(V)}_{\pi \gamma}(0)+F^{(NV)}_{\pi \gamma}(0)$
only half, and one can get the correct rate of the process
$\pi^0\to\gamma\gamma$ provided that the two basic contributions
$F^{(V)}_{\pi \gamma}(0)$ and $F^{(NV)}_{\pi \gamma}(0)$ are
considered simultaneously. By taking into account the PCAC
prediction~\cite{pcac}, $F_{\pi\gamma}(0)=1/(4\pi^2 f_\pi)$, one can
obtain the important constraint of the pion wavefunction, i.e.
\begin{equation}
\label{Bconstrain} \int^1_0 dx \Psi_{q\bar{q}}(x,{\bf
k}_{\perp}=0)=\frac{\sqrt{3}}{f_{\pi}}.
\end{equation}

Without loss of generality, we can assume that the pion wavefunction
depending on $\mathbf{k}_\perp$ through $k_\perp^2$ only, i.e.
$\Psi_{q\bar{q}}(x,\mathbf{k}_\perp)=\Psi_{q\bar{q}}(x,k_\perp^2)$
\footnote{The spin-space Wigner rotation might change this property
for the higher helicity components as shown in Ref.\cite{hms}. Since
the higher helicity components' contribution are highly suppressed
for the present case, we do not take this point into consideration
in the present paper.}. Then $F^{(V)}_{\pi \gamma}(Q^2)$
(Eq.(\ref{fflead})) can be simplified after doing the integration
over the azimuth angle as~\cite{musatov}
\begin{equation}\label{simplea}
F^{(V)}_{\pi \gamma}(Q^2)=\frac{1}{4\sqrt{3}\pi^2}\int_0^1\frac{d
x}{x Q^2}\int_0^{x^2 Q^2}\Psi_{q\bar{q}}(x,k_\perp^2) d k^2_\perp .
\end{equation}
Similarly, for the first derivative of $F^{(NV)}_{\pi \gamma}(Q^2)$
over $Q^2$, we have
\begin{equation}\label{derivative}
F^{(NV)'}_{\pi \gamma}(Q^2)|_{Q^2\to
0}=\frac{1}{8\sqrt{3}\pi^2}\left[\frac{\partial}{\partial
Q^2}\int_0^1\int_{0}^{x^2
Q^2}\left(\frac{\Psi_{q\bar{q}}(x,k_\perp^2)}{x^2 Q^2}\right)dx
dk_\perp^2\right]_{Q^2\to 0}.
\end{equation}
Furthermore, the leading twist-2 pion DA at the factorization scale
$\mu$ can be simplified as
\begin{equation}\label{phimu}
\phi_\pi(x,\mu^2)=\frac{\sqrt{3}}{8\pi^2 f_\pi}\int_0^{\mu^2}
\psi_{q\bar{q}}(x,k^2_\perp) dk^2_\perp .
\end{equation}

With the help of Eqs.(\ref{simplea},\ref{phimu}), $F^{(V)}_{\pi
\gamma}(Q^2)$ can be rewritten as
\begin{equation}
F^{(V)}_{\pi \gamma}(Q^2)=\frac{2f_\pi}{3Q^2}\int_0^1
\frac{dx}{x}\phi_\pi(x,x^2 Q^2).\label{simpv}
\end{equation}
Note that Eq.(\ref{simpv}) is different from Eq.(\ref{fLepage}) only
by replacing $\phi_\pi(x,Q^2)$ to $\phi_\pi(x,x^2 Q^2)$. It means
that the leading contribution to $F_{\pi\gamma}(Q^2)$ as shown in
Eq.(\ref{fflead}), which was given by keeping the
$k_\perp$-corrections in both the hard-scattering amplitude and the
pion wavefunction, can be equivalently obtained by setting the upper
limit for the integral of the pion DA to be $[\mu^2=x^2 Q^2]$. The
$x$-dependent upper limit $[x^2 Q^2]$ affects
$F^{(V)}_{\pi\gamma}(Q^2)$ from the small to intermediate $Q^2$
region, and such effect will be more explicit for a wider pion DA,
such as the CZ (Chernyak-Zhitnitsky)-like model~\cite{cz} that
emphasizes the end-point region in a strong way, as has been
discussed in Ref.\cite{huang1}. In the literature, the pion DA is
usually expanded in Gegenbauer polynomial expansion as
\begin{equation}\label{qcde0}
\phi_\pi(x,\mu^2)=\phi_{as}(x)\cdot\left[1+\sum^\infty_{n=1}
a_{2n}(\mu^2)C_{2n}^{3/2}(\xi) \right],
\end{equation}
where $\xi=(2x-1)$, $C^{3/2}_n(\xi)$ are Gegenbauer polynomials and
$a_{2n}(\mu^2)$, the so called Gegenbauer moments, are hadronic
parameters that depend on the factorization scale $\mu$. The
Gegenbauer moments $a_{2n}(\mu^2)$ can be related to
$a_{2n}(\mu_0^2)$ with the help of QCD evolution, where $\mu_0$
stands for some fixed low energy scale. To leading logarithmic
accuracy, we have~\cite{melic,nlo}
\begin{equation}\label{qcde}
a_{2n}(\mu^2)=a_{2n}(\mu_0^2)\left(\frac{\alpha_s(\mu^2)}{\alpha_s(\mu_0^2)}
\right)^{\gamma_0^{(2n)}/(2\beta_0)},
\end{equation}
where $\beta_0=11-2n_f/3$,
$\alpha_s(Q^2)=4\pi/[\beta_0\ln(Q^2/\Lambda_{QCD}^2)]$ and the
one-loop anomalous dimension is
\begin{equation}
\gamma_0^{(2n)}=8C_F\left(\psi(2n+2)+\gamma_E-\frac{3}{4}
-\frac{1}{(2n+1)(2n+2)}\right).
\end{equation}

We need to know $\phi_\pi(x,x^2 Q^2)$ and $F^{(NV)}_{\pi
\gamma}(Q^2)$ to get the whole behavior of $F_{\pi \gamma}(Q^2)$,
i.e.
\begin{equation}\label{final0}
F_{\pi\gamma}(Q^2) =F^{(V)}_{\pi\gamma}(Q^2)+
F^{(NV)}_{\pi\gamma}(Q^2),
\end{equation}
where $F^{(V)}_{\pi\gamma}(Q^2)$ is determined by $\phi_\pi(x,x^2
Q^2)$. The $\phi_\pi(x,x^2 Q^2)$ depends on the behavior of
$\Psi_{q\bar{q}}(x,\mathbf{k}_\perp)$ and its Gegenbauer moments can
not be directly obtained from the QCD evolution equation
(\ref{qcde}), since $[x^2 Q^2]$ can be very small and then the
Landau ghost singularity in the running coupling $\alpha_s$ can not
be avoided. As for $F^{(NV)}_{\pi \gamma}(Q^2)$, Eq.(\ref{ffb0})
presents an expression only at $Q^2\sim0$ region, and it can not be
directly extended to the whole $Q^2$ region. Ref.\cite{musatov} did
an attempt to understand the higher $Q^2$ behavior of $F_{\pi
\gamma}(Q^2)$ within the QCD sum rule approach, i.e. they raised a
simple picture: the sum over the soft $\bar{q}G\cdots Gq$ Fock
components is dual to $q\bar{q}$-state generated by the local axial
vector current. Furthermore, they raised an `effective' two-body
pion wavefunction that includes all soft contributions from the
higher Fock states based on the QCD sum rule analysis and then
calculated $F_{\pi \gamma}(Q^2)$ within the pQCD approach. Here we
will not adopt such an `effective' pion wavefunction to do the
calculation, since we plan to extract some information for the
leading Fock-state wavefunction by comparing with the CLEO
experimental data.

In order to construct an expression of $F^{(NV)}_{\pi \gamma}(Q^2)$
in the whole $Q^2$ region, we require the following conditions at
least:

i) $F^{(NV)}_{\pi \gamma}(Q^2)|_{Q^2=0}$ should be given by
Eq.(\ref{ffb00}).

ii) $F^{(NV)'}_{\pi \gamma}(Q^2)|_{Q^2\to 0}=\partial F^{(NV)}_{\pi
\gamma}(Q^2)/\partial Q^2|_{Q^2\to0}$ should be derived from
Eq.(\ref{ffb0}).

iii) $\frac{F^{(NV)}_{\pi \gamma}(Q^2)}{F^{(V)}_{\pi
\gamma}(Q^2)}\to 0$, as $Q^2\to\infty$.

\noindent One can construct a phenomenological model for
$F^{(NV)}_{\pi \gamma}(Q^2)$ that satisfies the above three
requirements. It is natural to assume the following form
\begin{equation}\label{model}
F^{(NV)}_{\pi \gamma}(Q^2)=\frac{\alpha}{(1+Q^2/\kappa^2)^2},
\end{equation}
where $\kappa$ and $\alpha$ are two parameters that can be
determined by the above conditions (i,ii), i.e.
\begin{equation}
\alpha=\frac{1}{2}F_{\pi\gamma}(0)=\frac{1}{8\pi^2 f_\pi}
\end{equation}
and
\begin{equation}
\kappa=\sqrt{-\frac{F_{\pi\gamma}(0)}{\frac{\partial}{\partial
Q^2}F^{(NV)}_{\pi \gamma}(Q^2)|_{Q^2\to 0}}}.
\end{equation}
As for the phenomenological formula (\ref{model}), it is easy to
find that $F^{(NV)}_{\pi \gamma}(Q^2)$ will be suppressed by $1/Q^2$
to $F^{(V)}_{\pi \gamma}(Q^2)$ in the limit $Q^2\to\infty$. Such a
$1/Q^2$-suppression is reasonable, since the phenomenological
expression (\ref{model}) can be regarded as a summed up effect of
all the high twist structures of the pion wavefunction, even though
each higher twist structure is suppressed by at least $1/Q^4$
\cite{musatov}.

\section{calculated results with the model wavefunction}

\begin{figure}
\centering
\includegraphics[width=0.50\textwidth]{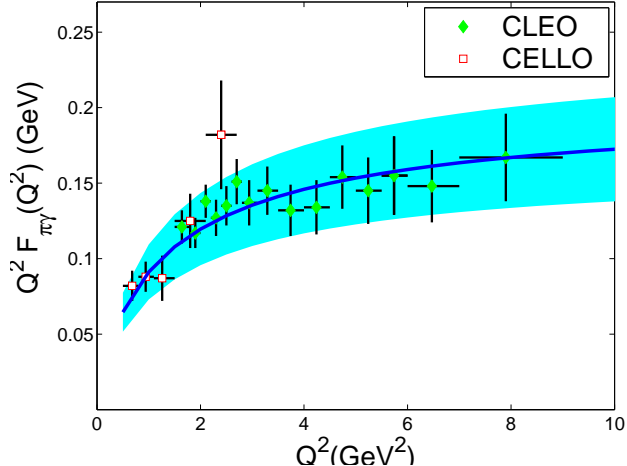}
\caption{The fitting curve (the solid line) for
$Q^2F_{\pi\gamma}(Q^2)$ from the CLEO and CELLO experimental
data~\cite{CLEO,CELLO}, where a shaded band shows its uncertainty of
$\pm 20\%$.} \label{piphoton}
\end{figure}

The CLEO collaboration has measured the $\gamma\gamma^*\to \pi^0$
form factor~\cite{CLEO}. In this experiment, one of the photons is
nearly on-shell and the other one is highly off-shell, with a
virtuality in the range $1.5$~GeV$^2$ - $9.2$~GeV$^2$~\cite{CLEO}.
There also exists older experimental results obtained by the CELLO
collaboration ~\cite{CELLO}. By comparing the theoretical prediction
with the experimental results, it provides us a chance to determine
a precise form for the leading Fock-state pion wavefunction. Similar
attempt to determine the pion DA has been done in literature
\cite{sch1,doro}, e.g. Ref.\cite{sch1} used the QCD light-cone sum
rule analysis of the CLEO data to obtain parameters of the pion DA.
In Fig.\ref{piphoton}, we show the fitting curve for
$Q^2F_{\pi\gamma}(Q^2)$ (derived by using the conventional
$\chi^2$-fitting method described in Ref.\cite{book} with slight
change to make the curve more smooth) from the CLEO and CELLO
experimental data, i.e. under the region of
$Q^2\in[0.5,10.0]$~GeV$^2$, $Q^2F_{\pi\gamma}(Q^2)\simeq [8.81\times
10^{-7}(Q^{'2})^5-4.78\times 10^{-5}(Q^{'2})^4+
9.96\times10^{-4}(Q^{'2})^3 -1.01\times10^{-2} (Q^{'2})^2
+5.29\times10^{-2} (Q^{'2}) + 4.48\times10^{-2}]$~GeV with the
dimensionless parameter $Q^{'2}=Q^2/{\rm GeV}^2$. Here the shaded
band shows its $\pm 20\%$ uncertainty \footnote{It is so chosen
since the sum of the statistical and systematic errors of the
experimental data is $\lesssim\pm 20\%$~\cite{CLEO,CELLO}.}. In
fact, most of the results given in literature, e.g. Refs.
\cite{huang1,ma1,job1,rus1,sch1,melic,musatov,kim}, are mainly
within such region. The shaded band (region) for
$Q^2F_{\pi\gamma}(Q^2)$ can be regarded as a constraint to determine
the pion wavefunction, i.e. the values of the parameters in the pion
wavefunction should make $Q^2F_{\pi \gamma}(Q^2)$ within the region
of the shaded band as shown in Fig.(\ref{piphoton}).

Now we are in position to calculate the pion-photon transition from
factor with the help of Eq.(\ref{final0}). As has been discussed in
the last section, we need to know the leading Fock-state pion
wavefunction $\Psi_{q\bar{q}}(x,\mathbf{k}_\perp)$ so as to derive
$\phi_\pi(x,x^2 Q^2)$ that is necessary for $F^{(V)}_{\pi
\gamma}(Q^2)$ and to derive the values of $\alpha$ and $\kappa$ for
$F^{(NV)}_{\pi \gamma}(Q^2)$. Several non-perturbative approaches
have been developed to provide the theoretical predictions for the
hadronic wavefunction. One useful way is to use the approximate
bound-state solution of a hadron in terms of the quark model as the
starting point for medeling the hadronic wavefunction. The
Brodsky-Huang-Lepage (BHL) prescription~\cite{bhl} for the hadronic
wavefunction is obtained in this way by connecting the equal-time
wavefunction in the rest frame and the wavefunction in the infinite
momentum frame. In the present paper, we shall adopt the revised LC
harmonic oscillator model as suggested in Ref.\cite{hms} to do our
calculation, which is constructed based on the BHL-prescription. As
discussed in the above section, the contribution from the higher
helicity states $\left(\lambda_1+\lambda_2=\pm 1\right)$ is highly
suppressed in comparison to that of the usual helicity state
$\left(\lambda_1+\lambda_2=0\right)$, so we only write down the form
of the pion wavefunction for the usual helicity state:
\begin{equation}\label{wave}
\Psi_{q\bar{q}}(x,{\bf k}_{\perp})=\varphi_{\mathrm{BHL}}(x,{\bf
k}_{\perp}) \chi^K(x,{\bf k}_{\perp})=A \exp\left[-\frac{{\bf
k}_{\perp}^2 +m^2}{8{\beta}^2x(1-x)}\right]\chi^K(x,{\bf
k}_{\perp}),
\end{equation}
with the normalization constant $A$, the harmonic scale $\beta$ and
the quark mass $m$ to be determined. The spin-space wavefunction
$\chi^K(x,{\bf k}_{\perp})$ can be written as \cite{hms}, $
\chi^K(x,{\bf k}_{\perp}) = m/\sqrt{m^2+k_\perp^2}$ with
$k_\perp=|{\bf k}_{\perp}|$. By taking the BHL-like wavefunction
(\ref{wave}), $F^{(V)}_{\pi \gamma}(Q^2)$ (Eq.(\ref{simplea})) can
be simplified as
\begin{equation}
F^{(V)}_{\pi \gamma}(Q^2)=\int_0^1 dx \left\{\frac{A m
\beta}{\sqrt{6}\pi^{3/2}Q^2} \sqrt{\frac{x'}{x}} \left({\rm
Erf}\left[\frac{\sqrt{m^2+x^2Q^2}} {2\beta\sqrt{2x x'}}\right]- {\rm
Erf}\left[\frac{\sqrt{m^2}} {2\beta\sqrt{2x x'}}\right]\right)
\right\}, \label{final3}
\end{equation}
where the error function ${\rm Erf}(x)$ is defined as
$\mathrm{Erf}(x)=\frac{2}{\sqrt{\pi}}\int_0^x e^{-t^2}dt$. And
similarly, for the limiting behaviors of $F^{(NV)}_{\pi
\gamma}(Q^2)$ that are necessary to determine the parameters
$\alpha$ and $\kappa$, we obtain
\begin{equation}
F^{(NV)}_{\pi \gamma}(Q^2)|_{Q^2\to 0}=\frac{A}{8\sqrt{3}\pi^2}
\int_0^1 \exp\left[-\frac{m^2}{8{\beta}^2x x'}\right] dx
\end{equation}
and
\begin{equation}
F^{(NV)'}_{\pi \gamma}(Q^2)|_{Q^2\to 0}=\frac{-A}{128\sqrt{3}m^2
\pi^2\beta^2} \int_0^1 \frac{x }{x'}(m^2+4x x' \beta^2)
\exp\left[-\frac{m^2}{8{\beta}^2x x'}\right] dx ,\label{radiusb}
\end{equation}
where $x'=1-x$.

\begin{figure}
\centering
\includegraphics[width=0.50\textwidth]{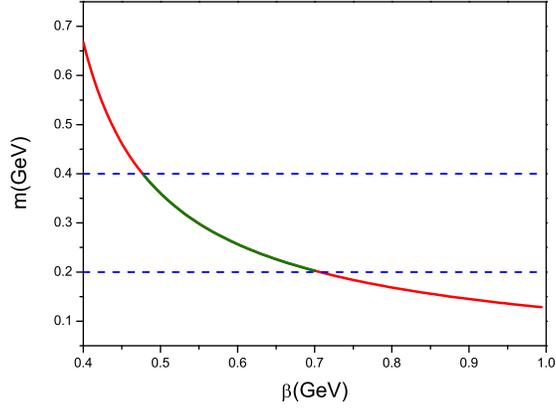}
\caption{The curve for the value of $m$ versus $\beta$, which shows
that if $m$ is in the reasonable region $[0.20\;{\rm GeV},
0.40\;{\rm GeV}]$ then $\beta\in[0.48\;{\rm GeV}, 0.70\; {\rm
GeV}]$.} \label{mbeta}
\end{figure}

The possible range for the parameters of the pion wavefunction
(\ref{wave}) can be derived by comparing our result of $Q^2F_{\pi
\gamma}(Q^2)$ with the experimental data. Further more, the model
wavefunction should satisfy the following two conventional
constraints:
\begin{itemize}
\item The pion wavefunction satisfies the
normalization that is derived from the $\pi\to\mu\nu$ process
~\cite{hms,bhl}:
\begin{equation}
\label{Aconstrain} \int^1_0 dx \int_{|\mathbf{k}_\perp|^2<\mu^2}
\frac{d^{2}{\bf k}_{\perp}}{16\pi^3}\Psi_{q\bar{q}}(x,{\bf
k}_{\perp}) =\frac{f_{\pi}}{2\sqrt{3}},
\end{equation}
where the cut $|\mathbf{k}_\perp|^2<\mu^2$ has been explicitly
included. Substituting the pion wavefunction (\ref{wave}) into
Eq.(\ref{Aconstrain}), we obtain
\begin{equation}
\int_0^1 \frac{A m \beta\sqrt{x(1-x)}}{4\sqrt{2}\pi^{3/2}}\left(
\mathrm{Erf}\left[\sqrt{\frac{m^2+\mu^2}{8\beta^2
x(1-x)}}\right]-\mathrm{Erf}\left[\sqrt{\frac{m^2}{8\beta^2
x(1-x)}}\right] \right)dx=\frac{f_{\pi}}{2\sqrt{3}} .\label{final1}
\end{equation}
As for the model wavefunction (\ref{wave}), it can be found that the
contribution from higher $|\mathbf{k}_\perp|$ region to the
wavefunction normalization drops down exponentially, e.g. by taking
reasonable values for the wavefunction parameters, the wavefunction
in the region of $|\mathbf{k}_\perp|> 1$~GeV only contributes about
$5\%$ to its total normalization, which changes to be less than
$1\%$ in the region of $|\mathbf{k}_\perp|> 2$~GeV. For clarity, we
take the factorization scale $\mu$ to be $\mu=\mu_0\simeq2$~GeV,
where $\mu_0$ stands for some hadronic scale that is of order ${\cal
O}(1~{\rm GeV})$ and the choice of $\mu_0\simeq2$~GeV is also close
to the virtuality of the photons in the central region of the
experimental data. By taking $1\;{\rm GeV}\leq\mu_0\leq 2\;{\rm
GeV}$ that is of order $\mu_0\sim{\cal O}(1~{\rm GeV})$, we find
that the following results will be slightly changed, especially for
the DA moments due to the fact that they satisfy the QCD evolution
equation (\ref{qcde}) within errors. A similar discussion on this
point can also be found in Ref.\cite{sch1}. Furthermore, one can
safely take $\mu=\infty$ to simplify the calculation due to the
smallness of the wavefunction in the region of $[|\mathbf{k}_\perp|>
\mu_0=2\;{\rm GeV}]$.

\item Another constraint, as shown in Eq.(\ref{Bconstrain}), for
the pion wavefunction can be derived from $\pi^0\rightarrow
\gamma\gamma$ decay amplitude~\cite{bhl}, which can be further
simplified as
\begin{equation}
\int_0^1 A \exp\left[-\frac{m^2}{8{\beta}^2
x(1-x)}\right]dx=\frac{\sqrt{3}} {f_{\pi}} . \label{final2}
\end{equation}
\end{itemize}

Solving Eqs.(\ref{final1},\ref{final2}) numerically, we obtain an
approximate relation for $m$ and $\beta$, i.e.
\begin{equation}\label{relationas}
6.00\frac{m\beta}{f_{\pi}^2}\cong
1.12\left(\frac{m}{\beta}+1.31\right) \left(\frac{m}{\beta}+5.47
\times 10^1\right),
\end{equation}
which shows that the value of $m$ is decreased with the increment of
$\beta$. More explicitly, Fig.(\ref{mbeta}) shows that $\beta$
should be within the region of $[0.48\;{\rm GeV}, 0.70\;{\rm GeV}]$
so as to restrict $m$ within the reasonable region of $[0.20\;{\rm
GeV}, 0.40\; {\rm GeV}]$.

Before doing the numerical calculation for $F_{\pi\gamma}(Q^2)$, we
note a naive interpolation formula for both perturbative and
non-perturbative regions has been proposed by Brodsky and Lepage
~\cite{lb}, which is similar to the monopole form derived from VMD
approach \cite{monopole}, i.e.
\begin{equation}\label{inter}
F^{BL}_{\pi \gamma}(Q^2)=\frac{1}{4 \pi^2 f_\pi
(1+Q^2/s_0)}\left(1-\frac{5}{3}\frac{\alpha_s(Q^2)}{\pi}\right)\;,\;\;
s_0=8\pi^2 f_\pi^2=0.67\; {\rm GeV}^2\sim m_\rho^2
\end{equation}
where the NLO perturbative contribution
$\left(-\frac{5}{3}\frac{\alpha_s(Q^2)}{\pi}\right)$ has been added
to the original result according to the suggestion of
Ref.\cite{musatov,brodsky2}.

\begin{figure}
\centering
\includegraphics[width=0.50\textwidth]{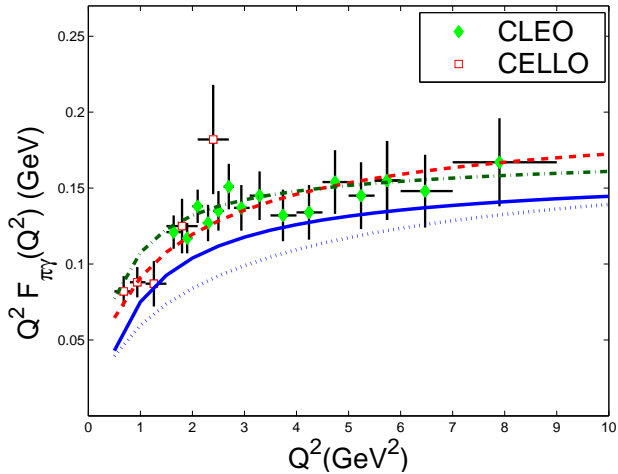}
\caption{$Q^2F_{\pi\gamma}(Q^2)$ that is defined in
Eq.(\ref{final0}) and is calculated with the model wavefunction
(\ref{wave}) for the case of $m=0.30$~GeV ($\beta=0.55$~GeV), which
is shown by a dash-dotted line. The dashed line is best fit for the
CLEO experimental data~\cite{CLEO,CELLO}, the solid line is from the
interpolation formula (\ref{inter}) and the dotted line is the
results for $Q^2F^{(V)}_{\pi\gamma}(Q^2)$ only.} \label{methodc}
\end{figure}

The conventional value for the constitute quark mass $m$ is around
$0.30$~GeV. In the present section, we concentrate our attention on
the case of $m=0.30$~GeV and we will study the uncertainty caused by
varying $m$ within a wider region of $[0.20\;{\rm GeV}, 0.40\;{\rm
GeV}]$ in the next section. We show the pion-photon transition form
factor $Q^2F_{\pi \gamma}(Q^2)$ with the model wavefunction
(\ref{wave}) for the case of $m=0.30$~GeV in Fig.(\ref{methodc}),
where the values for $\beta=0.55$~GeV and $A=25.6$~GeV$^{-1}$ can be
obtained by using Eqs.(\ref{Aconstrain},\ref{relationas}).
Fig.(\ref{methodc}) shows that by taking the model wavefunction
(\ref{wave}) with $m=0.30$~GeV, the predicted value of
$Q^2F_{\pi\gamma}(Q^2)$ agrees well with the experimental data. We
also show the leading Fock-state contribution
$Q^2F^{(V)}_{\pi\gamma}(Q^2)$ in Fig.(\ref{methodc}), which is drawn
in a dotted line and its value is lower than the experimental data
especially in low $Q^2$ region. As a comparison, we draw the curve
for the interpolation formula (\ref{inter}) in Fig.(\ref{methodc}),
where similar to Ref.\cite{cleo3}, we adopt the one loop
$\alpha_s$-running with $\Lambda^{N_f=3}_{QCD}=312$~MeV to do our
calculation. It shows that $Q^2F^{BL}_{\pi \gamma}(Q^2)$ agrees with
the data especially for the low $Q^2$ region, which is reasonable as
the effective value of $s_0$ in $F^{BL}_{\pi \gamma}(Q^2)$ is
determined by the known behavior of $Q^2\to 0$.

\section{discussion and comment}

\subsection{Information of the leading Fock state}

As shown in Fig.(\ref{methodc}), $F_{\pi \gamma}(Q^2)$ agrees well
with the experimental data by taking the leading Fock-state pion
wavefunction (\ref{wave}) with $m=0.30$~GeV. We also show the
leading Fock-state contribution $Q^2F^{(V)}_{\pi\gamma}(Q^2)$ in
Fig.(\ref{methodc}), which is lower than the experimental data in
low $Q^2$ region. This shows that one should take into account the
higher Fock states in small to intermediate $Q^2$ region. In fact,
it has been found that the leading Fock-state contribution
$F^{(V)}_{\pi \gamma}(Q^2)$ fail to reproduce the $Q^2=0$ value
corresponding to the axial anomaly~\cite{huang1,job1}, i.e. it gives
only half of what is needed to get the correct
$\pi^0\to\gamma\gamma$ rate~\cite{pcac}. And to make a compensation,
in Refs.\cite{ma1,musatov,rady2}, the leading Fock-state
contribution $F^{(V)}_{\pi \gamma}(Q^2)$ has been enhanced by
replacing the leading Fock-state wavefunction to an `effective'
valence quark wavefunction that is normalized to one. By taking the
`effective' pion wavefunction with the asymptotic-like DAs, the
authors found an agreement with the experimental data for
$F_{\pi\gamma}(Q^2)$~\cite{ma1,musatov,rady2}. However, such an
`effective' pion wavefunction is no longer the leading Fock-state
wavefunction itself and the probability of finding the leading Fock
state in pion should be less than one.

By substituting the pion wavefunction (\ref{wave}) into the pion
electro-magnetic form factor, one can obtain some useful
information, such as the probability of finding the leading Fock
state in pion $P_{q\bar{q}}$, the mean square transverse-momentum of
the leading Fock state $\langle\mathbf{k}_\perp^2\rangle_{q\bar{q}}$
and the charged mean-square-radius $\langle
r_{\pi^+}^2\rangle^{q\bar{q}}$. In
Refs.\cite{highhelicity,highhelicity2}, the authors have done such a
calculation within the LC pQCD approach. By adopting the formulae
derived in Ref.\cite{highhelicity} (Eqs.(24,25,27) there), one may
obtain:
\begin{equation}\label{pionef}
P_{q\bar{q}}=56\%, \;\langle\mathbf{k}_\perp^2\rangle_{q\bar{q}}
=(0.502\;{\rm GeV})^2, \; \langle r_{\pi^+}^2\rangle^{q\bar{q}}=
(0.33\;{\rm fm})^2 ,
\end{equation}
where we have taken $m=0.30$~GeV. All the values in
Eq.(\ref{pionef}) are summed results for all the helicity states
$\lambda_1+\lambda_2=(0,\pm1)$ of the leading Fock state.
Eq.(\ref{pionef}) shows that the value of $\langle
r_{\pi^+}^2\rangle^{q\bar{q}}$ is smaller than the pion charged
radius $\langle r^2\rangle^{\pi^+}_{expt}=(0.671\pm 0.008\; {\rm
fm})^2$~\cite{radius}, but it is close to the value as suggested in
Refs.\cite{highhelicity,pole}. Such small $\langle
r_{\pi^+}^2\rangle^{q\bar{q}}$ for the leading Fock-state
wavefunction is reasonable, since the probability of leading Fock
state $P_{q\bar{q}}$ is only $56\%$, which confirms the necessity of
taking the higher Fock-states into consideration to give full
estimation of the pion electromagnetic form factor/pion-photon
transition form factor, especially for lower $Q^2$ regions.

\subsection{Moments of the DA from the CLEO data}

\begin{figure}
\centering
\includegraphics[width=0.50\textwidth]{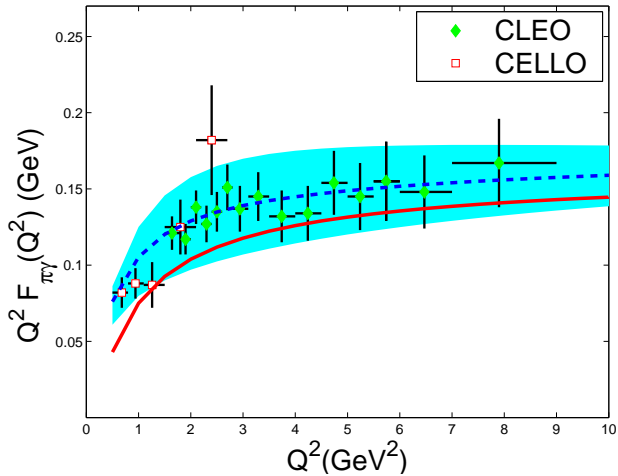}
\caption{$Q^2F_{\pi\gamma}(Q^2)$ with the phenomenological model
(\ref{model}) for $F^{(NV)}_{\pi \gamma}(Q^2)$, with the upper edge
of the band for $m=0.40$~GeV and the lower edge of the band for
$m=0.20$~GeV. The dashed line is the best fit of
$Q^2F_{\pi\gamma}(Q^2)$ for wavefunction (\ref{wave}) with
$m=0.30$~GeV. As a comparison, the interpolation formula
(\ref{inter}) is shown by a solid line.} \label{compare}
\end{figure}

In this subsection, we take a wider region for $m$, i.e.
$m=0.30^{+0.10}_{-0.10}$~GeV, to study the Gegenbauer moments of
pion DA. Under such region for  $m$, we show the value of
$Q^2F_{\pi\gamma}(Q^2)$ in Fig.(\ref{compare}). The value of
$Q^2F_{\pi\gamma}(Q^2)$ will increase with the increment of $m$ and
the uncertainty caused by varying $m$ within the region of
$[0.20\;{\rm GeV}, 0.40\;{\rm GeV}]$ is about $\pm 20\%$. By
comparing Fig.(\ref{compare}) with Fig.(\ref{piphoton}), it can be
found that $m\in(0.20\;{\rm GeV}, 0.40\;{\rm GeV})$ is a reasonable
region for $Q^2F_{\pi\gamma}(Q^2)$, which is in agreement with the
experimental data in the whole $Q^2$ region.

The pion DA in Eq.(\ref{DAdefinition}) can be defined through the
matrix elements of local operators between the physical pion and the
vacuum state \cite{rady1,brodsky5}. In the light-cone framework, by
choosing the frame in which $P=(1,0,\mathbf{0}_\perp)$ that has been
adopted in Sec.II to calculate the pion-photon transition form
factor, its explicit form is given by the following equation
\cite{brodsky4}:
\begin{equation}\label{defphi1}
\phi(x_i,Q^2)=\int\frac{dz^-}{2\pi}e^{i(x_1-x_2)z^-/2}\times
\left\langle0\left|\bar{u}(-z)\frac{\gamma^+\gamma_5}{2\sqrt{2}}
[-z,z] d(z)\right|\pi(P)\right\rangle^{(Q^2)}_{z^+=z_\perp=0},
\end{equation}
where the factorization scale is taken to be $Q^2$ and the
path-order factor $[-z,z]={\cal P}\exp\left(ig_s\int_{-1}^1 ds A^+(z
s)z^{-}/2\right)$ stands for the conventional Wilson line connecting
the points $-z$ and $z$. The line integral vanishes in the
light-cone gauge due to $A^{+}=0$. The behavior of $\phi(x_i,Q^{2})$
for the large $Q^{2}$ is dominated by the behavior of the operator
$T(\bar{u}(-z)d(z))$ for $z^{2}={\cal O}(1/Q^2)$. One can apply the
usual OPE and only gauge invariant operator that is actually
contributes to the matrix element in the light-cone gauge. Therefore
only the $q\bar{q}$ component is required in the above definition
for the leading power behavior \cite{brodsky5}. Also the
corresponding higher twist operators of the wave function matrix
elements are suppressed at the short distance by power of $1/Q^{2}$.
Thus, the DA $\phi(x_i,Q^{2})$ defined in Eq.(\ref{defphi1}) is the
amplitude for finding the $q\bar{q}$ component that is collinear up
to the scale $Q^{2}$. On the other hand, the matrix element
$\langle0|\bar{u}\gamma_\mu\gamma_5 d|\pi(P)\rangle $ can be
expressed by the following expansion \cite{pball}
\begin{eqnarray}
&&\langle0|\bar{u}(-z)\gamma_\mu\gamma_5
[-z,z]d(z)|\pi(P)\rangle^{(Q^2)}=\nonumber\\
&&\quad\quad\quad i\sqrt{2}f_\pi P_\mu \int_0^1 du \,
e^{i(2u-1)P\cdot z}\phi_\pi(u,Q^2) +\frac{i}{\sqrt{2}} m_\pi^2
\frac{z_\mu}{P\cdot z}\int_0^1 du e^{i(2u-1)P\cdot
z}g_\pi(u,Q^2),\label{defphi2}
\end{eqnarray}
where $\phi_\pi$ is the leading twist-2 DA, $g_\pi$ is the twist-4
DA. Under the light-cone gauge, the DA $\phi(x_i, Q^{2})$ defined in
Eq.(\ref{defphi1}) corresponds to the leading twist-2 DA $\phi_\pi$
defined in Eq.(\ref{defphi2}), except for an overall normalization
factor. Both definitions are consistent with each other.

We can derive the leading twist-2 pion DA that is in the usual
helicity $(\lambda_1+\lambda_2=0)$ from Eq.(\ref{wave}), i.e.
\begin{equation}\label{phipion}
\phi_\pi(x,\mu_0^2)=\frac{A m
\beta\sqrt{3}\sqrt{x(1-x)}}{2\sqrt{2}f_{\pi}\pi^{3/2}}\left(
\mathrm{Erf}\left[\sqrt{\frac{m^2+\mu_0^2}{8\beta^2
x(1-x)}}\right]-\mathrm{Erf}\left[\sqrt{\frac{m^2}{8\beta^2
x(1-x)}}\right] \right),
\end{equation}
where the fixed low energy scale $\mu_0$ is taken to be $2$~GeV and
$\phi_\pi(x,\mu_0^2)$ satisfies the normalization $\int_0^1
\phi_\pi(x,\mu_0^2)dx=1$. By expanding Eq.(\ref{phipion}) into the
Gegenbauer polynomials, for $m\in[0.20\;{\rm GeV}, 0.40{\rm GeV}]$,
we obtain
\begin{eqnarray}
&&a_2(\mu_0^2)=0.002^{+0.063}_{-0.054},\;
a_4(\mu_0^2)=-0.022_{-0.012}^{+0.026},\nonumber\\
&&a_6(\mu_0^2)=-0.014_{+0.000}^{+0.009},\;
a_8(\mu_0^2)=-0.006_{-0.001}^{+0.003},\; \cdots , \label{a2a4}
\end{eqnarray}
where the center value is for taking $m\simeq 0.30$~GeV that best
fits the CLEO experimental data, i.e. it has the minimum
$\chi^2$-value, and the ellipsis stands for higher Gegenbauer
moments. For the values of $a_{2n}(\mu^2)$ in other factorization
scales, they can be derived by QCD evolution e.g. Eq.(\ref{qcde}).
Eq.(\ref{a2a4}) shows: A) the leading twist-2 pion wavefunction
(\ref{wave}) is asymptotic-like, since $a_{2n}\; (n\geq 1)$ are much
smaller than $a_0\equiv 1$. More explicitly, the first inverse
moment of the pion DA at energy scale $\mu_0$, $\int_0^1
dx\phi_\pi(x,\mu_0^2)/x=3(1+a_2+a_4+a_6+a_8)\in(2.71,3.20)$, which
is near the same value as for the asymptotic wavefunction with
$a_{2n}=0\; (n\geq 1)$. Such a conclusion for pion wavefunction
agrees with that of Ref.\cite{kroll} and also agrees with a recent
study that is based on the nonlocal chiral-quark model from the
instanton vacuum~\cite{new}. B) $a_{2}$, $a_4$ will increase with
the decrement of $m$, and $a_2\geq0$ if $m\leq 0.30$~GeV. C) the
absolute values of $a_4$, $a_6$ and $a_8$ are comparable to $a_2$
for bigger $m$ (e.g. $m\sim 0.30$~GeV); but they are suppressed to
$a_2$ about one order for smaller $m$. The value of the Gegenbauer
moments have been studied in various processes, cf.
Refs.\cite{new,cleo2,cleo3,a2a41,a2a42,sch1,a2a44,a2lattice1,a2lattice2,a2lattice3,a2a45}.
The lattice result of Ref.\cite{a2lattice3} prefers a narrower DA
with $a_2(1\;{\rm GeV}^2)=0.07(1)$, while the lattice
results~\cite{a2lattice1,a2lattice2} prefer wider DA, i.e. they
obtain $a_2(1\;{\rm GeV}^2)=0.38\pm0.23^{+0.11}_{-0.06}$ and
$a_2(1\;{\rm GeV}^2)=0.364\pm0.126$ respectively. However as argued
in Ref.\cite{topicee}, the accuracy of the lattice results needs to
be further improved, e.g. the results in Ref.\cite{a2lattice2} for
the second moment $\langle \xi^2\rangle$ are obtained for the
``pion" with the masses $\mu_\pi>550$~MeV and then extrapolated to
the chiral limit $\mu_\pi\to 0$. These references favor a positive
value for $a_2(1\;{\rm GeV}^2)$ and the most recent one is done by
Ref.\cite{a2a42}, which shows that $a_2(1\;{\rm GeV}^2)=0.19\pm
0.19$ and $a_4(1\;{\rm GeV}^2)\geq -0.07$ by analyzing the leptonic
mass spectrum of $B\to\pi l\nu$. Here, the range of $m$ should be
reduced to $m\in[0.20\;{\rm GeV}, 0.30\;{\rm GeV}]$ if we require
$a_2(1\;{\rm GeV}^2)\geq 0$ for the pion DA. Or inversely, we have
$a_2(1\;{\rm GeV}^2)\in[0,0.08]$ with the help of the QCD evolution
equation (\ref{qcde}) for $m\in[0.20\;{\rm GeV}, 0.30\;{\rm GeV}]$.

\subsection{Comparison with the broad wavefunction}

One typical broad wavefunction is described by the CZ-like
wavefunction, which can not be excluded by the pion-photon
transition form factor~\cite{huang1} although it is disfavored by
the pion structure function at $x\to1$~\cite{hms}.

\begin{figure}
\centering
\includegraphics[width=0.50\textwidth]{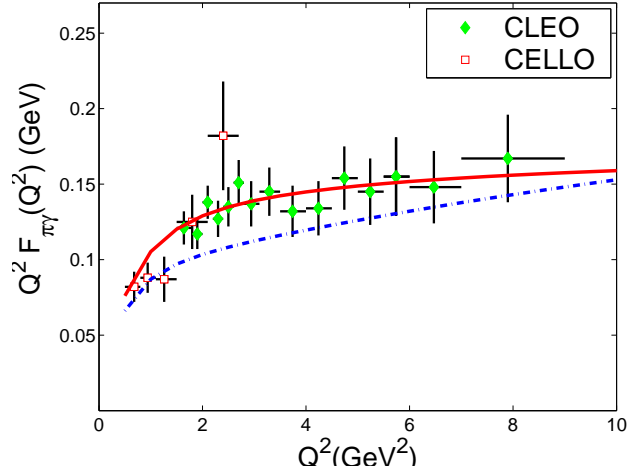}
\caption{Comparison of $Q^2F_{\pi\gamma}(Q^2)$ that is derived from
two different types of wavefunctions, i.e. the BHL-like wavefunction
(the solid line) and the CZ-like wavefunction (the dash-dotted
line), under the condition of $m=0.30$~GeV.} \label{asczcompare}
\end{figure}

We take the CZ-like wavefunction as
\begin{equation}\label{cz}
\Psi_{q\bar{q}}^{CZ}(x,{\bf k}_{\perp})=A(1-2x)^2
\exp\left[-\frac{{\bf k}_{\perp}^2
+m^2}{8{\beta}^2x(1-x)}\right]\chi^K(x,{\bf k}_{\perp}),
\end{equation}
where an extra factor $(1-2x)^2$ is introduced into the pion
wavefunction (\ref{wave})~\cite{cz}. Following a similar procedure,
we find that the relation between $m$ and $\beta$ changes to
\begin{equation}
6.00\frac{m\beta}{f_{\pi}^2}\cong
1.24\left(\frac{m}{\beta}+2.12\right) \left(\frac{m}{\beta}+4.58
\times 10^1\right).
\end{equation}
Similarly, for $m=0.30$~GeV, we have
\begin{equation}\label{pionCVef}
P_{q\bar{q}}=73\%, \;\langle\mathbf{k}_\perp^2\rangle_{q\bar{q}}
=(0.496\;{\rm GeV})^2, \; \langle r_{\pi^+}^2\rangle^{q\bar{q}}=
(0.45\;{\rm fm})^2 ,
\end{equation}
where all the values are summed results for all the helicity states
$\lambda_1+\lambda_2=(0,\pm1)$ of the leading Fock state. One may
observe that the second Gegenbauer moment $a_{2}$ is always dominant
over other higher Gegenbauer moments for the CZ-like DA. And for the
first inverse moment of the CZ-like pion DA at energy scale $\mu_0$,
we obtain $\int_0^1 dx\phi_\pi^{CZ}(x,\mu_0^2)/x =4.69$.

In Fig.(\ref{asczcompare}), we make a comparison of
$Q^2F_{\pi\gamma}(Q^2)$ that is derived from two different types of
wavefunctions, i.e. the BHL-like wavefunction and the CZ-like
wavefunction, under the same value of $m=0.30$~GeV. Both the
BHL-like wavefunction and the CZ-like wavefunction lead to
$Q^2F_{\pi\gamma}(Q^2)$ within the possible region of the
experimental data as shown in Fig.(\ref{piphoton}). However, the
value of $Q^2F_{\pi\gamma}(Q^2)$ derived from the BHL-like
wavefunction is better than that of the CZ-like wavefunction. One
may observe that the value of $Q^2F_{\pi\gamma}(Q^2)$ caused by the
CZ-like wavefunction shall increase with the increment of $m$, so
the CZ-like model can give a better result for
$Q^2F_{\pi\gamma}(Q^2)$ in comparison to the experimental data only
by taking a bigger value for $m$, e.g. at least, $m=0.40$~GeV.

As a summary, the main differences for the BHL-like wavefunction and
the CZ-like wavefunction are listed in the following:
\begin{itemize}
\item By comparing with the experimental data for
$Q^2F_{\pi\gamma}(Q^2)$, one may observe that
$m=0.30_{-0.10}^{+0.10}$~GeV is a reasonable region for the BHL-like
wavefunction, where the best fit to the experimental data is
achieved when $m\approx 0.30$~GeV; while for the case of the CZ-like
wavefunction, such region is shifted to a higher one, i.e.
$m=0.40_{-0.10}^{+0.10}$~GeV, and the best fit to the experimental
data is achieved as $m\approx 0.40$~GeV, which is somewhat bigger
than the conventional value for the constitute quark mass of pion.

\item The difference among the Gegenbauer moments e.g.
$a_2(\mu_0^2)$ are big due to the different behavior of the two
models. Under the condition of $m=0.30$~GeV, the two Gegenbauer
moments $a_2(\mu_0^2)=0.002$ and $a_4(\mu_0^2)=-0.022$ for the
BHL-like model (\ref{model}); while for the CZ-like model
(\ref{cz}), $a_2(\mu_0^2)=0.678$ and $a_4(\mu_0^2)=-0.024$.

\item The first inverse moments are different. Under the condition
of $m=0.30$~GeV, for the case of the BHL-like model (\ref{model}),
$\int_0^1 dx\phi_\pi^{BHL}(x,\mu_0^2)/x=2.88$, which is close to
that of the asymptotic DA; while for the case of the CZ-like model
(\ref{cz}), $\int_0^1 dx\phi_\pi^{CZ}(x,\mu_0^2)/x =4.69$, which is
close to that of the original CZ-model~\cite{cz}.
\end{itemize}

\begin{figure}
\centering
\includegraphics[width=0.50\textwidth]{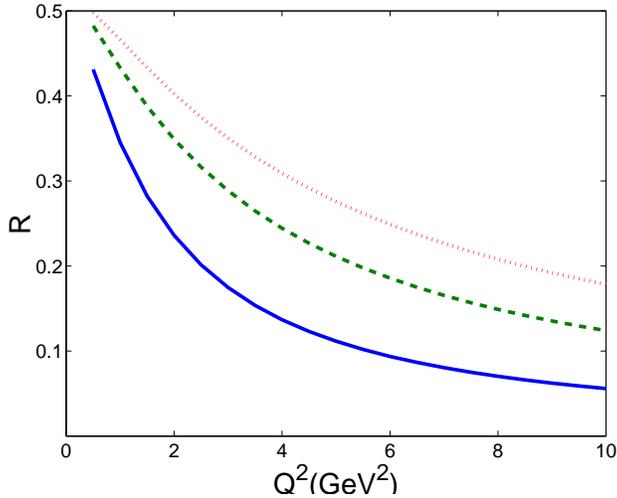}
\caption{The value of $R=\frac{F^{(NV)}_{\pi\gamma}(Q^2)}
{F^{(V)}_{\pi\gamma}(Q^2)+F^{(NV)}_{\pi\gamma}(Q^2)}$ versus $Q^2$
for the BHL-like wavefunction, whose value increases with the
increment of $m$. The solid, dashed and dotted lines are for
$m=0.20GeV$, $m=0.30GeV$ and $m=0.40GeV$, respectively.}
\label{ratio}
\end{figure}

In Ref.\cite{topicee}, it was shown that by taking the asymptotic DA
and considering the suppression from the NLO contribution, the value
of $Q^2F_{\pi\gamma}(Q^2)$ at $Q^2=8.0\;{\rm GeV}^2$ is somewhat
smaller than the experimental value $(16.7\pm2.5\pm0.4)\cdot
10^{-2}\;{\rm GeV}$, i.e. $Q^2F_{\pi\gamma}(Q^2)|_{Q^2=8.0\;{\rm
GeV}^2}\simeq 0.115$~GeV. However it should be pointed out that
Ref.\cite{topicee} is only a simplified analysis, since the effects
caused by the $k_T$-dependence in both the pion wavefunction and the
hard scattering amplitude, and the contribution from
$F^{(NV)}_{\pi\gamma}(Q^2)$ have not been taking into consideration.
As shown in Fig.(\ref{ratio}), even though
$F^{(NV)}_{\pi\gamma}(Q^2)$ is suppressed to
$F^{(V)}_{\pi\gamma}(Q^2)$ in high $Q^2$ region, it has sizable
contribution in the intermediate $Q^2$ region, e.g. it is about
$12\%$ at $Q^2=8.0\;{\rm GeV}^2$ for the case of $m=0.30$~GeV.

\subsection{A simple discussion of the NLO correction}

Up to NLO, by taking the square of the factorization scale $\mu_F^2$
to be $Q^2$, the pion-photon transition form factor
$F_{\pi\gamma}(Q^2)$ can be schematically written as
\begin{equation}\label{addmodel}
F_{\pi\gamma}(Q^2)
=F^{(V)}_{\pi\gamma}(Q^2)\left(1-\delta\cdot\frac{\alpha_s(Q^2)}{\pi}\right)
+ F^{(NV)}_{\pi\gamma}(Q^2),
\end{equation}
where $\delta$ is a parameter that is to be determined by the
behavior of the pion wavefunction/DA and the detail form of the hard
scattering amplitude. It is a natural choice to take $\mu_F^2=Q^2$,
which directly eliminates the $\mu_F$-dependence in the terms that
determine how much of the collinear term is absorbed into the
distribution amplitude \cite{a2a44,melic}. In the literature, the
value of $\delta$ has been estimated to be: $\delta=\frac{5}{3}$ for
asymptotic DA \cite{musatov,brodsky2,melic} and
$\delta=\frac{49}{108}$ for DA in CZ form \cite{musatov}. It should
be noted that to be consistent with our present calculation
technology, we need a full NLO result, in which all the effects
caused by the transverse-momentum dependence in the hard-scattering
amplitude and the wavefunction and by the Sudakov factor should be
fully taken into consideration. However such a full NLO calculation
is not available at the present. The value of $\delta$ will be
decreased by considering the transverse-momentum dependence in the
hard-scattering amplitude and the wavefunction (a naive discussion
for this point can be found in Refs.\cite{musatov,kim}). Since our
model wavefunction is close to asymptotic-like one, we simply take
$\delta=\frac{5}{3}$ to do our discussion.

Under the condition of $\delta=\frac{5}{3}$, it can be found that
the best fit of $Q^2F_{\pi\gamma}(Q^2)$ for the case of the BHL-like
wavefunction (\ref{wave}) is at $m\simeq 0.32$~GeV. If taking
$m=0.32\pm0.10$~GeV, we obtain $a_2(\mu_0^2)=-0.02^{+0.07}_{-0.08}$,
where $\mu_0\simeq 2$~GeV. Furthermore, if taking $a_2(1\;{\rm
GeV}^2)>0$ as an extra constraint, we find that $m$ must be in the
region of $[0.23,0.30]$~GeV. Or inversely, we have $a_2(1\;{\rm
GeV}^2)\in[0,0.06]$ for $m\in[0.23\;{\rm GeV}, 0.30\;{\rm GeV}]$.

\section{Summary}

In this paper, we have given a careful analysis of the pion-photon
transition form factor $F_{\pi\gamma}(Q^2)$ involving the transverse
momentum corrections with the present CLEO experimental data, in
which the contributions beyond the leading Fock state have been
taken into consideration. As is well-known, the leading Fock-state
contribution dominates the pion-photon transition from factor
$F_{\pi\gamma}(Q^2)$ for large $Q^2$ region and it gives only half
contribution to $F_{\pi\gamma}(0)$ as one extends it to $Q^2=0$. One
should consider the higher Fock states' contribution to
$F_{\pi\gamma}(0)$ at the present experimental $Q^2$ region. We have
constructed a phenomenological expression to estimate the
contributions beyond the leading Fock state based on the limiting
behavior of $F^{(NV)}_{\pi\gamma}(Q^2)$ at $Q^2\to 0$. The
calculated results favor the asymptotic-like behavior by comparing
the predictions from different type of the model wavefunctions with
the experimental data.

On the other hand, the present CLEO data provides the important
information of the pion DA as one has a complete expression for the
pion-photon transition form factor relates two photons with one
pion. Our expression for $F_{\pi\gamma}(Q^2)$ only involves a single
pion DA. Thus, comparing the calculated results of $F_{\pi
\gamma}(Q^2)$ by taking the BHL-like pion wavefunction with the CLEO
data one can extract some useful information of the pionic leading
twist-2 DA. Our analysis shows that (1) the probability of finding
the leading Fock state in the pion is less than one, i.e.
$P_{q\bar{q}}=56\%$ and $\langle r_{\pi^+}^2\rangle^{q\bar{q}}=
(0.33\;{\rm fm})^2$ with $m = 0.30$~GeV. This means that the leading
Fock state is more compact in the pion and it is necessary to take
the higher Fock states into account to give full estimation of the
pion-photon transition form factor and other exclusive processes.
(2) under the region of $m\in[0.20\;{\rm GeV}, 0.40\;{\rm GeV}]$, we
have the DA moments: $a_2(\mu_0^2)=0.002^{+0.063}_{-0.054}$,
$a_4(\mu_0^2)=-0.022_{-0.012}^{+0.026}$ and all of higher moments.
Such result is helpful to understand other exclusive processes
involving the pion.

\begin{center}
{\bf ACKNOWLEDGEMENTS}
\end{center}

This work was supported in part by the Natural Science Foundation of
China (NSFC). X.-G. Wu thanks the support from the China
Postdoctoral Science Foundation. \\

\end{document}